\documentstyle[aaspp4,flushrt]{article}

\lefthead{Poli et al.}
\righthead{ }

\begin{document}

\title{THE EVOLUTION OF THE GALAXY SIZES IN THE NTT DEEP FIELD:\\ A COMPARISON
WITH CDM MODELS}
\author{F. Poli$^1$, E. Giallongo$^1$, N. Menci$^1$, S. D'Odorico$^2$, A. Fontana$^1$
}

\bigskip

\affil{$^1$ Osservatorio Astronomico di Roma, via dell'Osservatorio, I-00040
Monteporzio, Italy\\
\noindent
$^2$ European Southern Observatory, Karl Schwarzschild Strasse 2,
D-85748 Garching, Germany\\
}

\begin{abstract}

The sizes of the field galaxies with $I\leq 25$ have been measured in
the NTT Deep Field. Intrinsic sizes have been obtained after
deconvolution of the PSF with a multigaussian method. The reliability
of the method has been tested using both simulated data and HST
observations of the same field. The distribution of the half light
radii is peaked at $r_{hl}\simeq 0.3$ arcsec, in good agreement with
that derived from HST images at the same magnitude. An approximate
morphological classification has been obtained using the asymmetry and
concentration parameters. The intrinsic sizes of the galaxies are
shown as a function of their redshifts and absolute magnitudes
using photometric redshifts derived from the multicolor catalog. While
the brighter galaxies with morphological parameters typical of the
normal spirals show a flat distribution in the range $r_{d}=1-6$ kpc,
the fainter population at $0.4<z<0.8$ dominates at small sizes. To
explore the significance of this behaviour, an analytical rendition of
the standard CDM model for the disc size evolution has been computed.
The model showing the best fit to the local luminosity function and
the Tully-Fisher relation is able to reproduce at intermediate
redshifts a size distribution in general agreement with the
observations, although it tends to underestimate the number of
galaxies fainter than $M_B\sim -19$ with disk sizes $r_d\sim 1-2$ kpc.

\end{abstract}

\keywords {galaxies: fundamental parameters -- galaxies: evolution -- 
galaxies: formation}

\section{INTRODUCTION}

Deep multicolor surveys from HST and ground based telescopes coupled
with deep spectroscopy on ten meter class telescopes (Keck, VLT) are
leading to a remarkable progress in our understanding of the galaxy
formation and evolution.
In particular, there has been great improvement in the knowledge
of the epoch of star formation. The first determinations of the global
star formation history suggested a peak at $z\approx 1.5$ with a
further decline from $z\simeq 2$ to $z\approx 4$ (Madau et al. 1996;
Connolly et al. 1997).  In this scenario, more than 80\% of the stars
would have been formed recently at $z<2$. 

Such a picture would be in rough agreement with the cold dark matter
(CDM) models of galaxy formation, which predict the star formation to
be a gradual process governed by gas cooling and Supernovae feedback.
The timescale of the process is strongly related to the rate at which
the mass is hierarchically assembled to form larger galaxies from
smaller structures (White \& Frenk 1991; Cole et al. 1994; Kauffmann,
White \& Guiderdoni 1993; Baugh et al. 1998).

However, the above scenario is presently challenged by recent
indications that the observed decline at $z>2$ is affected by
incompleteness in the color selection and by dust extinction. In fact,
more recent observational evidence limited to the optically
selected galaxies, points towards a flat SFR for $z>2$ (Giallongo et
al. 1998, Pascarelle et al. 1998, Steidel et al. 1999).

A more precise test of the galaxy formation scenarios involves the
study of the size and morphological distribution of galaxies
with cosmic time, since this is related to the main feature of the
hierarchical models, namely, the growth of mass and sizes of typical
condensations.  Indeed, although the global star formation history is a
crucial issue in galaxy formation, it is an integrated quantity; even
a flat SFR could be consistent with the CDM picture by tuning the
parameters describing the gas cooling and the Supernovae feedback in
galaxies.

The first observational investigations of the differential
evolution of the galaxy population over the redshift range $0<z<1$
were carried out by Lilly et al. (1995), Cowie et al. (1996) and Ellis
et al. (1996) by means of spectroscopic redshift surveys. The observed
decrease with the cosmic time of the blue star-forming galaxy
population showing strong emission lines has been further investigated
using the Hubble Space Telescope.  The decline is a mixture of stellar
evolution in normal spirals and strong evolution in low luminosity
irregulars (Schade et al. 1995, Brinchmann et al. 1998, Lilly et
al. 1998) whose local counterparts now comprise the faint end of the
local luminosity function (Zucca et al. 1997).

Subsequently, the study of the Hubble Deep Field (Williams et
al. 1996) has resulted in a strong improvement in the knowledge of the
galaxy morphology at high redshift (Abraham et al. 1996, Odewahn
et al. 1996, Driver et al. 1998) showing a high fraction of faint
irregular galaxies.  Most of these galaxies have small sizes resulting
in a physical extent of the order of $3-5$ kpc at $z\sim 1$
(Glazebrook et al. 1998). Such small sizes, if compared with the present
day galaxy morphology included in the Hubble sequence, could imply
physical growth and merging occurring with increasing cosmic time
(Bouwens, Broadhurst and Silk 1997), in qualitative agreement with the
hierarchical clustering picture.  On the other hand, the indications
for a slow evolution of the spiral population constituted by the
presence of large, relatively bright spirals at $z\sim 0.6$ (Schade et
al. 1995) are difficult to fit in the picture of the CDM-like
models. Indeed, the dissipative formation of galactic disks in CDM
halos predicted by the hierarchical clustering scenario produces
spirals whose size strongly decreases as the formation redshift
increases. For this reason, gas cooling in the CDM halos has to be
delayed to low redshifts ($z<1$).  The comparison becomes more
stringent for the lower luminosity spirals where the predicted sizes
are very small at high redshifts.

A wide field survey of faint galaxies observed with the Hubble Space
Telescope would be the ideal sample to clarify the issue and to
provide a differential and quantitative test to CDM galaxy formation
theories. However deep photometry from HST can only be obtained on
small areas. On the contrary, wide field surveys of faint galaxies can
be obtained using ground-based instrumentation but at lower resolution
(FWHM$\sim 0.5-1$ arcsec).

In this paper we have exploited the deep photometric information of the NTT
Deep Field (Arnouts et al. 1999) obtained under condition of subarcsec
seeing at the ESO NTT telescope to derive morphological information
after appropriate seeing deconvolution (Sect. 2). The reliability of
the deconvolution technique has been tested by means of numerical
simulations and particularly by checking on somewhat shallower HST images
of the same galaxy field retrived from the HST archive. The results
show that it is possible to derive the radial intensity profiles of
galaxies with $I\leq 25$ from ground-based imaging with FWHM$<1$
arcsec (Sect. 3).

The derived intrinsic angular sizes are then converted into physical
sizes adopting reliable photometric redshifts estimated from the
UBVRIJK multicolor distribution of each galaxy in the catalog.  The
galaxy morphological distribution in redshift and luminosity is shown
in Sect.~4 and is compared with that predicted by hierachical
clustering models for the disk formation in Sect.~5. The conclusions
are presented in Sect.~6.

\section{THE GALAXY CATALOG}

The galaxy catalog has been obtained from the NTT Deep Field (NDF)
(Arnouts et al. 1999).  The optical observations were obtained in
the Johnson BV, and Gunn R,I broadband filters with the $1k\times 1k$
SUSI camera on the ESO New Technology Telescope. The resulting field
of view is $\sim 5$ arcmin$^2$. 

The $3 \sigma$ magnitude limits are $B=27.5$, $V=27$, $R=26.5$,
$I=26.5$, the average seeing was $\sim 0.8$ arcsec (see Arnouts et
al. 1999 for further details).

Deep ultraviolet images have been added with the SUSI-2 camera at the
NTT reaching a 3$\sigma$ upper limit of $U=26$.  Also deep near
infrared observations have been added with the SOFI instrument at the
NTT reaching 3$\sigma$ magnitude limits of $J\sim 23.5$ and
$K^{\prime}\sim 21.7$.

The galaxy catalog has been extracted from the sum of the BVRI
weighted frames to detect objects at very faint level in the four
bands (Arnouts et al. 1999). The object detection and photometry have
been performed with the Sextractor image analysis package (Bertin \&
Arnouts 1996). Total magnitudes in the UBVRIJK bands were computed
from isophotal magnitudes for galaxies brighter than I=23.25 and from
aperture magnitudes in 2.2 arcsec corrected to 5 arcsec for fainter
objects. Colors are computed in a 2 arcsec diameter aperture in the
same sky area in all the bands after correction for the different
seeing conditions. Details on the data reduction and photometry are
fully discussed in Fontana et al. (1999).

To each galaxy in the UBVRIJK catalog, a photometric redshift estimate
has been assigned with the same best fitting procedure applied in the
contiguous BR1202-0725 field (Giallongo et al. 1998).  This has been
obtained through a comparison of the observed colors with those
predicted by spectral synthesis (GISSEL) models (Bruzual \& Charlot
1993) including UV absorption by the intergalactic medium and dust
reddening. The catalog of the photometric redshifts is presented in
Fontana et al. (1999). The resulting typical redshift accuracy is
$\Delta z \sim 0.1$ up to $z\sim 1.5$ and $\Delta z\sim 0.15$ at
larger redshifts.

The redshift distribution in the NTT Deep Field is shown in Fig.~1 for
galaxies with $I\leq 25$. As derived in other fields of similar
depth, the bulk of the galaxies present are at intermediate redshifts
$z\sim 0.5-0.8$ with a tail in the distribution up to $z\sim 4$.

The shape of the optical counts in this field suggests negligible
correction for incompleteness at $I=25$ (Arnouts et al. 1999)
thus the morphological study described in the next sections should be
considered complete to this magnitude limit.

\section{THE RADIAL INTENSITY PROFILES OF FAINT GALAXIES}

The estimate of the intrinsic radial intensity profiles derived from
deep ground based images of faint galaxies is a complex task. Indeed, 
HST images of galaxies with $I\sim 25$ have shown that most of them
have typical sizes corresponding to half light radii $r_{hl}\sim 0.3$
arcsec which are smaller than the typical HWHM$\sim 0.5$ arcsec
present in ground based deep images.

Accurate and reliable deconvolution techniques must be applied to
obtain any statistical information about the intrinsic half light
radii of faint galaxies observed from the ground. Any deconvolution
technique depends of course on the accurate knowledge of the
instrumental profile (PSF) and on the intrinsic shape assumed for the
galaxies.  An accurate reconstruction of the instrumental profile is
generally done through an average of the stellar profiles with high
s/n ratio in images with good oversampling.

Concerning the intrinsic shape of the galaxies, instead of assuming a
specified profile, we adopted the multigaussian deconvolution analysis
studied in detail by Bendinelli (1991) and also recently applied to
faint elliptical galaxies in the HDF by Fasano et al. (1998).

For each object detected in the sum of the BVRI weighted frames, a
radial intensity profile was obtained, sampling the light into
elliptical annuli with axes proportional to the object
intensity-weighted second order moments.  Following Fasano et
al. (1998) we chose to carry out the analysis on a circularly
symmetrized profile, where the radial distance along the semi-major
axis $a_{n}$ is $r_{n}=\frac{a_{n}}{\sqrt{\epsilon}}$, where
$\epsilon$ is the Sextractor elongation parameter.

A multi gaussian expansion of these symmetrized profiles has been
developed as described below. If $I(r)$ is the true,
deconvolved profile of the object, it is possible to obtain a simple
but effective approximation of it:
\begin{equation}
I(r)=L_{T}\sum_{j=1}^{M}\frac{b_{j}}{2\pi s_{j}^{2}}
     exp\left(-\frac{r^{2}}{2s_{j}^{2}}\right)
\label{eq:prof}
\end{equation}
where $L_{T}$ is the total luminosity of the source, $s_{j}$ and
$b_{j}$ the spread parameter and the weight of each gaussian,
respectively.  We adopt $M=6$ as a compromise between the need of
sampling the profile with the maximum allowed number of gaussians and
the need of performing a well-posed fit. Weighting parameters are
obviously not independent: their sum must be equal to one.

In an analogous way an expansion of the normalized point spread
function in the frame was carried out, so that it was possible to have
an analytical solution of the convolution integral:

\begin{equation}
I_{conv}(r)=
L_{T}
\sum_{j=1}^{M}\sum_{i=1}^{N}
\frac{a_i b_j}{2\pi( s_{j}^{2}+\sigma_{i}^{2})}
\cdot
\exp{\left[
-\frac
{r^{2}}
{2\left(
s_{j}^{2}+\sigma_{i}^{2}
\right)}
\right]
}
\label{eq:mge}
\end{equation}
where the terms with index ``i'' refer to the multi gaussian expansion
of the PSF. Once the PSF expansion is known, it is possible to fit
Eq.~\ref{eq:mge} to the observed points, obtaining the spread and
weighting parameters $s_j$, $b_j$ of the deconvolved profile.

In order to verify the reliability of the deconvolution technique in our
particular case, we have performed two checks. First of all we have
produced a set of simulations specifically designed to reproduce
typical conditions in our sample. Second, we have compared, for a
bright galaxy subsample, intrinsic half light radii derived from the I
band galaxy profiles taken at the NTT with that derived from a relatively
deep image in the I band taken at the WFPC2 on HST.

In the first check, intensity profiles have been reproduced as in the
observed NTT deep field assuming the same pixel sampling and the same
average seeing (FWHM$\sim 0.8$ arcsec).  Assuming an intrinsic
exponential profile, a series of synthetic images were constructed
using different half light radii ranging from $r_{hl}=0.1$'' to
$r_{hl}=0.9$'' with a step of 0.1''. An average of 25 random objects
were computed for each radius, assuming a total magnitude of
$I\simeq 23.5$ and of $I\simeq 24.7$ which are typical values for a
bright and a faint object in our sample.  In this way the s/n ratio is
constant in the simulations.  In order to reproduce the average seeing
conditions in the NDF, the background subtracted image of a bright star
was selected in the field. Its normalized profile was then
convolved with the synthetic images of the disk galaxies.  The
convolved two dimensional profiles were randomly inserted in
regions of the NTT deep field far from very bright objects to
reproduce with the appropriate pixel size and noise levels the
observed NTT galaxies. Finally, the multigaussian deconvolution
technique was applied to the synthetic data. We first notice that
we have no selection bias against galaxies with large size
($r_{hl}\sim 1$ arcsec) and low surface brightness down to $I\simeq
24.7$ since all the synthetic objects were detected.

The results are shown in Fig.~2, where the error bars represent the
dispersion around the mean due to noise in the background subtraction. A
good match between the intrinsic and measured half light radii has
been obtained up to $r_{hl}\sim 0.7$ arcsec. For larger values, a
slight underestimate in the measured values appears particularly for the
faint subsample. In any case it can be seen that, even for the faint
galaxies with $I\sim 25$, the overall correlation between intrinsic
and measured half-light radii is preserved in such a way that an
intrinsically large, faint object, e.g. with $r_{hl}\sim 0.7''$, can not
be detected as a small sized one, e.g. with $r_{hl}\sim 0.1''$.  The
simulation shows that the fraction of small size galaxies observed in
the NDF is real and is not due to intrinsically larger objects which
have been shrunk by noise effects.

For the brighter galaxies we had the opportunity to check directly the
reliability of our deconvolution technique comparing our I band image
photometry of the NTT deep field with the I band WFPC2 image
photometry taken from the HST archive.

We have applied to the WFPC2 galaxies the same deconvolution technique
using an average PSF derived from the available stellar profiles
in the field.  Since the WFPC2 image is not as deep as the NTT I band
image and the overlap between the two images is limited to less than
half of the field, the comparison was limited to a few relatively
bright objects with $I<24.5$. In Fig.~3 the deconvolved half light radii
derived from ground based and HST I band images are compared. The
agreement is very good with a rms difference of 0.05. A similar
agreement for the galaxy scale lengths has been found by Schade et
al. (1996) for a sample of CFRS galaxies two mag brighter,
$I_{AB}<22.5$. Thus, although HST images allow more detailed
morphological studies with a decomposition of the intensity profiles
into more components (e.g. bulge plus disk decomposition), the
statistical information is limited by the small field of view. Large
statistical information on the galaxy $r_{hl}$ from wide field ground
based images can be used to constrain the models of galaxy formation.
As a first step, we have applied this technique to the multicolor NTT
deep field. The results are discussed in the next Section.

\section{THE DISTRIBUTION OF THE GALAXY SIZES IN LUMINOSITY AND REDSHIFT}

The distribution of the half light radii $r_{hl}$ for all the galaxies
in the multicolor catalog with $I<25$ is shown in Fig.~4. 
The median value of the distribution is $r_{hl}\simeq 0.3$'',
which confirms previous findings from the HST medium deep survey
(MDS). In fact, the derived distribution is very similar to that of
the MDS (Roche et al. 1998). This is an important result since
it has been obtained by means of ground based observations.

Assuming that all the galaxies are characterized by a disk with a given
scale length $r_d=r_{hl}/1.68$, we have computed the linear size for
each galaxy as a function of redshift using the color estimated
redshifts as discussed in section 2.  The resulting luminosity-size
relation is shown in Fig.~5 in two redshift intervals. Uncertainties
in the redshift estimates $\Delta z \sim 0.1$ produce uncertainties in
the sizes and luminosities $\sim 5$\% and  $\sim 30$\%, respectively.

Fig.~5 shows a concentration of faint galaxies with $r_d<3$ kpc that
is present at $M_B>-19$. This is more evident in the $0.4<z<0.8$
interval where the galaxy distribution extends down to $M_B\sim -16$.

The histogram of the size distribution is shown in Fig.~6 in the same
redshift intervals where a peak of low-luminosity dwarf galaxies
appears for $r_{d}\leq 2$ kpc.

However the assumption of a disk-like structure for the majority of
the faint galaxies can no longer be valid because of the presence of a
noticeable fraction of dwarf galaxies with irregular
morphology. Although it is not clear if the images of the high $z$
galaxies seen at progressively shorter wavelengths can produce a
spuriously increasing number of irregular galaxies, we have attemped to
exclude galaxies with irregular morphology following the method
described in Abraham et al. (1996).  With this method, galaxies are
separated into three classes (irregulars, spirals, S0-ellipticals)
according to the values of their asymmetry and concentration
parameters. The asymmetry parameter is defined as:
\begin{equation}
\displaystyle
A=\frac{\frac{1}{2}
        \sum
       |{I_{i,j}-I_{i,j}^{180}}|
         }
       {\sum I_{i,j}} - k
\label{eq:AC1}
\end{equation} 
where $I$ and  $I^{180}$ are the background-subtracted intensity  of the 
original image and the same after 180-degree rotation around
the baricenter, respectively. Here the sum is extended to the isophotal
detection area of every source, and $k$ takes account of the 
bias introduced in A by the underlying portion of the sky with the same
area and shape as the object. For the concentration parameter we have:
\begin{equation}
\displaystyle
C_{\alpha}=\frac{\sum_{i,j\epsilon S_{\alpha}}I_{i,j}}
           {\sum_{i,j\epsilon S}I_{i,j}}
\label{eq:AC2}
\end{equation} 
where $S$ refers to a circular aperture with a radius $r$ so that 
it has the same number of pixels as the isophotal detection area
of the object. In an analogous way $S_{\alpha}$ is an inner circular area
with radius $r_{\alpha}=\alpha r$. 
We chose to adopt $\alpha = 0.4$ instead of $\alpha = 0.3$ because we 
found that this larger value takes account of the seeing blurring 
of the sources in the NDF compared to the HST high resolution images.    
 The distribution in the
asymmetry-concentration plane shown in Fig.~7 is in good agreement
with the Abraham et al. (1996) distribution derived from the Hubble
Deep Field galaxies of comparable magnitude.

In Fig.~7 regular spiral galaxies are distributed in the central
region delimited by the two straight lines defined in Abraham et
al. (1996). Their distribution in the luminosity-size plane is also
shown by filled symbols in Fig.~5. We have reproduced in Fig.~8 the
size distribution of the normal spirals in the same redshift interval
as in Fig.~6.  It can be seen that even reducing the sample to the
more regular spirals, the size distribution in Fig.~8 is very similar
in shape to that of the whole sample shown in Fig.~6 with a peak at
very small sizes $r_d\sim 1.5$ kpc. This is expected since the
distribution of disk-dominated systems in the luminosity-size plane
(filled circles in Fig.~5) follows the distribution of the whole
sample.

This behaviour has interesting consequences for cosmological scenarios
of galaxies formation.

\section{EVOLUTION OF GALACTIC DISKS IN CDM COSMOLOGY}

The size distribution of different spectral types as a function of
redshift is in principle a powerful tool to constrain models of galaxy
formation in hierarchical CDM cosmogonies.

Indeed, in these models the galaxy formation proceeds through
subsequent merging of small units into larger ones, corresponding to a
hierarchical growth of the typical mass of virialized
condensations. Since these models predict galaxies in the range
$1<z<3$ to be still assembling, the corresponding size growth should
be detectable by present-day surveys.

Quantitatively, to compare the predictions of the CDM models with deep
imaging observations, one needs to express the outcomes of the
hierarchical clustering picture in terms of observable properties,
i.e., luminosities and sizes. The first goal is usually pursued
through the use of semi-analytical models (White \& Frenk 1991; Cole
et al. 1994; Kauffmann, White \& Guiderdoni 1993; Baugh et al. 1998),
which are able to predict luminosity functions and Tully-Fisher
relations in reasonable agreement with existing observations. As for
the galaxy sizes, quantitative attempts to relate them to the mass
growth of cosmic structures have been done only for the regular spiral
population (Mo, Mao \& White 1998).

To compare with observations we have to connect the galaxy sizes with
their luminous properties. To this aim, we first used the Mo et
al. (1998) model to connect the scale length of any exponential disk
$R_d$ to the dark matter (DM hereafter) circular velocity $v_c$; then
we used the approach of semi-analytical models to connect $v_c$ to the
luminous properties of galaxies.  The predicted relations among
luminosity, size and redshift of the field spirals have then been used
to derive statistical distributions which have been compared with the
observed ones.

\subsection{Size-circular velocity relation}

In the framework of present theories of galaxy formation the size of
the disk $R_d$ can be related to the circular velocity of the disk
($v_d$) and of the DM ($v_{c}$) halo, assuming that the mass
and angular momentum of the baryonic material which settles into the
disk are fixed fractions -- $m_d$ and $j_d$ -- of the DM mass $m$
and angular momentum $J$ (see Mo, Mao \& White 1998). The latter can be 
expressed in dimensionless units 
$\lambda\equiv J/(|E|^{1/2}/G\,m^{5/2})$, whose distribution $p(\lambda)$ is
known from several N-body experiments (see Warren et al. 1992; Cole \&
Lacey 1996; Steinmetz \& Bartelmann 1995).

For a galaxy DM density profile of the form given by Navarro, Frenk \&
White (1996, NFW hereafter), it is found that
\begin{equation} 
R_d={1\over \sqrt{2}}\,\Bigg({j_d\over m_d}\Bigg)\,\lambda\,{v_c\over 10\,H(z)}
\,f_c(c)^{-1/2}\,f_R(\lambda,c,m_d,j_d)~,
\end{equation}
where $H(z)$ is the Hubble constant at redshift $z$ and
$c=r_s/r_{200}$ is the concentration factor expressed in terms of the
scale radius $r_s$ in the NFW distribution and of the virial radius
$r_{200}$.  The function $f_c$ is related to the total energy of the
NFW profile while $f_R$ includes also the gravitational effect of the
disk in the density distribution. Fitting formulas for the last two
functions are provided by Mo et al. (1998).  

Note that in the above model the disk contributes to the
rotation velocity, increasing by more than 20 \% the value of
$v_c$ given by the DM halo. We have taken this effect into account
in our computations. The power spectrum of the initial DM
density perturbation (from which the DM halos are thought to be formed
via gravitational instability) enters the computation through the
concentration factor $c$.

According to the above discussion,  and following the lines in Mo et
al. (1998), we have computed the $R_d-v_d$ and the $R_d-v_c$ relations
for different CDM perturbation spectra. In order to compare the
results with the outcome of the data analysis we have converted the
$R_d-v_d$ (or $R_d-v_c$) relationships into $R_d-M_B$
as described in the next section.

\subsection{Size-magnitude relation using a model for galaxy formation} 

To implement the present understanding of the process of galaxy
formation and evolution -- and thus to associate magnitutes to the DM
galaxy masses -- we propose an analytical rendition of the
semi-analytical approaches introduced by Kauffman, White \& Guiderdoni
(1993) and by Cole et al. (1994) and subsequently developed by several
authors (see Somerville \& Primack 1998).  In these models the DM
halos form through the gravitational instability of initial random
density fluctuations whose power spectrum depends on the nature of the
DM. The baryons within the DM halo, heated to the virial temperature
of the halo, are assumed to be distributed in three phases: a cold
phase where the gas is radiatively cooled and star formation may
occur, a condensed phase (stars) and a hot phase (gas re-heated by
Supernovae explosions). The hierarchical merging of the halos
(followed by the possible merging of the galaxies contained in the
halos) and the luminosity evolution due to the evolution of the
stellar population are also included in the model.

The interplay among these phases and the details of our implementation
of the hierarchical models of galaxy formations are given in Appendix
A. Following the lines in the literature, the parameters regulating
the star formation have been chosen so as to match the local observed
properties of the galaxies, i.e., the I-band Tully-Fisher relation and
the B-band luminosity function. However, since the disk velocity $v_d$
is $\sim 20\%$ higher than that of the DM, matching the predictions of
the model with both the above observables is critical.  E.g., setting
the star formation parameters to maximize the match to the observed
Tully-Fisher relation for a ``Milky Way'' galaxy with $v_c$=220 km/s,
yields an excess of bright galaxies in the local luminosity
function (as noted by Somerville \& Primack 1998). A further and
complementary constraint is provided by the faint end of the
luminosity function; minimizing the mismatch between the observed and
predicted Tully-Fisher relation at faint magnitudes leads to a
luminosity function too steep at the faint end.
 
For these reasons, we have selected the set of free parameters that
minimizes the discrepancy of the model results with respect to both
the local Tully-Fisher relation
and the local B-band luminosity function. For the angular momentum
parameters (determining the $v_c/v_d$ ratio and thus affecting the
predicted Tully-Fisher relation) we have obtained $j_d=m_d=0.3$. For
the star formation parameters we have obtained the set reported in
Appendix A.  For the cosmology we have adopted a tilted CDM cosmogony
(normalized to COBE with $\Omega_0=1$ and $H_0=50$ km s$^{-1}$
Mpc$^{-1}$); the predictions of semi-analytical models with such a
power spectrum are very close to the mixed (hot+cold) DM models, which
are in fair agreement with the data for a wide set of observations
(see Primack et al. 1995).  With this recipe we have obtained a
reasonable agreement of the model prediction with the local luminosity
function and Tully-Fisher relation (Fig.~9). The growth of the disk
size with cosmic time, a characteristic prediction of CDM models, is
shown in fig. 10. The size distributions, plotted at different $z$ for
two ranges of absolute magnitudes, are computed convolving eq. 5 over
the full distribution $p(\lambda)$ of the halo angular momentum
$\lambda$ that a galaxy with given $R_d$ and $v_d$ can have at any z.
We have taken for $p(\lambda)$ the expression given in Mo et
al. (1998).

\subsection{Results and comparison with the observations} 

In Fig.~11 we show how the galaxy sizes are distributed in absolute
magnitude in four redshift intervals.  The dots are the observed data
while the shaded regions correspond to the CDM predictions. The latter
depend on the value of the dimensionless angular momentum $\lambda$
which is distributed according to the function $p(\lambda)$ in the
interval $0.025<\lambda<0.1$, the extreme values being the 10\% and
90\% points of $p(\lambda)$, respectively. The solid line corresponds
to $\lambda=0.05$, i.e., to the 50\% point of the distribution
$p(\lambda)$.

Note that at bright ($M_B<-19$) magnitudes the observed points
are approximately equally distributed above and below the
solid line, and fall in the region allowed by the CDM model. In fact,
for the bright galaxies (where a measurement of the disk
rotation velocity has been possible) the model by Mo et al. (1998),
relating size and circular velocity, is in agreement with the
observations; since our model Tully-Fisher relation relating $v_d$ to
$M_B$ is also in agreement with observations (see Fig.~9), the
matching between our predicted and observed distributions provides a
consistency check of the theoretical approach used to derive the
galaxy properties in CDM models.  However, at fainter magnitudes
($M_B>-19$), the observed sizes tend to occupy systematically the
small size region below the 50\% locus of the angular momentum
distribution.

To investigate more quantitatively this point, the differential size
distribution of galaxies with $I<25$ is shown in Fig.~8, for different
redshift bins. Here the model results include the full $p(\lambda)$
distribution to take account of all the values of angular momentum
that a galaxy with given $R_d$ and $v_d$ can have at a given $z$. To
compare the model with the observed data in our sample, we have also
introduced a magnitude limit at $I<25$. This limit has the effect of
selecting brighter (and hence larger) objects at higher $z$. This
effect tends to balance the decrease with $z$ of the disk size
predicted by CDM models (as shown in fig. 10 at given absolute
magnitude), resulting in an apparent lack of evolution with $z$ of the
distributions predicted by the CDM model (see Fig.~8).

The comparison with the data shows a general agreement, although an
excess of small-size objects is evident with respect to the CDM
predictions.  This points towards a galaxy size evolution from high to
low $z$ slower than predicted by CDM models, with the persistence of
small size objects dominating in number at faint magnitudes and at
intermediate to low redshifts.  The differential distributions in Fig.~8
are plotted in terms of absolute numbers in our 5 arcmin$^2$ field, so
that the excess at small sizes over the CDM predictions is not
affected by any possible incompleteness at the large/bright end of the
distribution, which could be present at lower $z$ in our limited
sized field.

We also note that the excess is robust with respect to any possible
systematic errors in the selection of disk galaxies using the
concentration-asymmetry plane shown in Fig.~7. Indeed, when the whole
sample is considered (including also bulge and irregular systems) the
resulting distributions shown in Fig.~6 are very similar to the
distributions of the ``spiral'' sample shown in Fig.~8. In particular,
the excess with respect to the CDM predictions at small sizes is still
present.

\section{CONCLUSIONS}

The size distribution in a ground based sample of faint galaxies to
$I=25$ is peaked at $r_{hl}\simeq 0.3$ arcsec, in good agreement with
that found at similar magnitudes in the HST medium deep survey.

A morphological selection based on the asymmetry and concentration
parameters gives a distribution similar to the one found by Abraham et al.
(1996) from HST data. From this distribution a sample of normal
spirals has been selected. Photometric redshifts have been obtained
for this sample using the available multicolor catalog.

A clear distinction in the morphological properties of galaxies
fainter or bighter than $M_B\simeq -19$ has been found, especially in
the $0.4<z<0.8$ redshift interval.  In particular the fainter galaxies
show a marked concentration toward small ($r_d< 3$ kpc) sizes (see
Fig.~5).

The comparison of the observed galaxy sizes with those predicted by
hierarchical clustering CDM models at intermediate redshifts ($z\sim
0.5-1$) indicates an excess of faint, small-size galaxies with respect to
the model predictions (see Figs.~6, 8, 11). A better agreement is
found at brighter ($M_B<-19$) magnitudes.

Different reasons could be responsible for this excess:

a) The {\it ad hoc} assumption that all the disks have masses and
angular momenta that are a fixed fraction of those of their DM
counterparts ($j_d\approx m_d=$ const.) is incorrect for small
galaxies.  In fact, numerical N-body simulations indicate that the gas
loses most of its angular momentum to the DM during galaxy assembly
(Navarro \& Benz 1991; Navarro \& White 1994; Navarro \& Steinmetz
1997), although the effect of Supernovae feedback (not included in the
simulation) is still unclear. In this case, however, the problem of
explaining why the assumption $j_d\approx m_d\approx 0.3$ works for
more massive galaxies should be solved.

b) The faint-end of the galaxy luminosity distribution evolves slower
than predicted by hierarchical models. The observed population of
small galaxies corresponds to the turn-off in the galaxy luminosity
function at the faint end, and those galaxies constitute a population
evolving in a physically different way from the body of the
distribution due, for example, to the lack of aggregation of objects
of small circular velocity. This would break the self-similarity of
the hierarchical clustering evolution at small galactic scales.

c) The luminosity of the satellite small galaxies within a given DM
halo does not take into account star bursting phenomena due to merging
interactions within the halo. For a fixed galaxy luminosity this
implies an excess of small size objects.

d) The evolution of the number of galaxies within a halo is poorly
described by semianalytical models. The two body aggregation processes
in fact could produce a strong number evolution resulting in an excess
of small objects as the redshift increases.

These results have to be checked on a larger sample where the
morphological parameters are measured on high resolution images. The
VLT telescopes could be ideal in this respect in terms of field of
view, photometric depth and optical quality.  Moreover, a caution
should be emphasized when we associate the galaxy morphological
properties derived from the observations to the unknown dynamical
properties of the faint galaxies. All the previous arguments apply
only if these faint objects are sustained by rotation.  Rotational
curves of galaxies with $I\sim 25$ can not be measured with the
current technology on 8m class telescopes.  Thus, the statistical
approach presented in this paper represents a first step to include
the galaxy morphology in the CDM models as a new constraint for galaxy
formation and evolution.

\acknowledgments 
We thank the referee Dr. S. Odewahn for his helpful comments. We are
grateful to S. Zaggia for providing HST archive images and A. Renzini
for discussions. The ultraviolet observations of the NTT Deep Field
were performed in SUSI-2 guaranteed time of the Observatory of Rome in
the framework of the ESO-Rome Observatory agreement for this
instrument.

\begin{appendix}

\section{The Model of Galaxy Formation}

The model proceeds through the following basic steps: 

{\bf a)} We define a time-circular velocity grid, with grid size
$\Delta t$ and $\Delta v_c$. Since a given DM halo may contain several
galaxies, we denote with $V_c$ the circular velocity of the halos
(whose DM mass is $M$) and with $v_c$ that of the galaxies; the latter
represents the depth of the galactic potential wells corresponding to
a mass $m$.

The number density $N(V_c,t(z_i))$ of Dark Matter (DM) halos with
circular velocity $V_c$ present at the initial redshift $z_i$ is taken
from the Press \& Schechter expression. Initially, we associate to
each DM halo a galaxy whose circular velocity $v_c$ is the same of the
halo.  To each $V_c$, a halo mass $M=V_c^3/[10\,G\,H(z)]$ and a virial
temperature $T_v$ are associated. The initial gas content associated
to all the halos corresponds to the universal baryonic density
$\Omega_b$. The initial stellar content is taken to be 0.  For each
circular velocity, the baryonic content is divided into a cold gas
phase (the one able to cool within the halo survival time $\tau_l$),
with mass $m_c$, into a star phase, with mass $m_*$, and the
complementary hot phase, with mass $m_h$.

{\bf b)} For each galaxy circular velocity $v_c$, we compute the variations
of the mass of the gas components and of the stellar content due to
merging of DM halos, possibly followed by galaxy merging.
At each time step, we first compute i) the probability $P(V_c',V_c,t)$
that a halo at time $t$ with velocity $V_c'$ is included in a halo
with velocity $V_c$ at time $t+\Delta t$, and ii) the associated halo
merging rate ${\partial P(V_c),t|V_c)/\partial V_c\partial t}$: these
are computed in the framework of the so-called Extended Press
\& Schechter Theory (EPST, see Bond et al. 1991; White \& Frenk 1991;
Bower 1991; Lacey \& Cole 1993).

Then we compute the number of galaxies $N_g(v_c,V_c,t)$ with circular
velocity $v_c$ in a halo with velocity $V_c$, implementing the
canonical recipes of semi-analytical models. According to these,
during halo merging, a galaxy contained in one of the parent halos
contributes to enrich the dominant galaxy of the common halo (of
velocity $V_c$) if its coalescence time $\tau_m (v_c)$ is less than
the halo survival time $\tau_l(V_c)$ (predicted in the EPST). In the
remaining case ($\tau_m>\tau_l$) the galaxy retains its identity. 

Thus, for galaxies with $v_c=V_c$, 
the positive increment in the number of galaxies of given $v_c$ in
halos with $V_c$ is linked to the number increment of DM halos with
$V_c$ [whose contribution from progenitors $V_c'$ is 
$N(V_c',t)\,\partial^2P(V_c',t|V_c)/\partial V_c'\partial t$] through the
probability $prob\big[\tau_m(v_c')<\tau_l (V_c)\big]$ of forming one 
dominant galaxy merging 
those contained in the parent halos; for galaxies with $v_c\neq V_c$ the 
positive increment is linked to the complementary probability
$1-prob\big[\tau_m(v_c')<\tau_l (V_c)\big]$. 
The overall negative
increment is due to disappearence of satellite galaxies of velocity
$v_c$ in halos $V_c$, when the latter merge to forme larger halos $V_c'$. 
Thus
\begin{eqnarray}
\lefteqn{N_g(v_c,V_c,t+\Delta t)   = N_g(v_c,V_c,t) \nonumber} \\
& &+\Delta t\,\delta(v_c,V_c)\int_0^{v_c}dv_c'\,\int_{v_c'}^{V_c}dV_c'
\,N(V_c',t)\,{\partial P(V_c',t|V_c)\over \partial V_c'\partial t}\,
{N_g(v_c',V_c')\over N_{gT}(V_c')}\,prob\big[\tau_m(v_c')<\tau_l (V_c)\big]
\nonumber \\
& &+ \Delta t\,\int_{v_c}^{V_c}dV_c'
\,N(V_c',t)\,{\partial P(V_c',t|V_c)\over \partial V_c'\partial t}\,
{N_g(v_c,V_c')\over N_{gT}(V_c')}\,
\Big\{1-prob\big[\tau_m(v_c')<\tau_l (V_c)\big]\Big\}\nonumber \\
& & -\Delta t\,\int_{V_c}^{\infty}\,dV_c'
{\partial P(V_c,t|V_c')\over \partial V_c\partial t}
\,N_g(v_c,V_c)~, 
\end{eqnarray}
where $\delta(v_c,V_c)=1$ for $v_c=V_c$ and 0 elsewhere. 

$N_{gT}(V_c)=\int dv_c'\,N_g(v_c',V_c,t)$ is the total number of
galaxies in the halo $V_c'$.  The merging time of the galaxies is
parametrized (in terms of the DM masses $M(V_c)$ and $M'(V_c')$) as
$\tau_m= \tau_o(M'/M)^{\alpha_{merg}}$ according to Cole et
al. (1994), while the probability $prob\big[\tau_m(v_c')<\tau_l
(V_c)\big]$ for a DM halo of velocty $V_c$ to have a survival time
$\tau_l$ larger than a given value has been computed by Lacey \& Cole
(1993) (see their eq. 2.21) in the framework of the EPST. The initial
condition is such that in each halo there is only one galaxy (i.e.,
$N_g(v_c,V_c,t_0)=N(V_c,t_0)$ for $v_c=V_c$ and 0 elsewhere).  From
the ratio $N_g(v_c,V_c,t)/N(V_c,t)$ we compute the probability
$f(v_c,V_c,t)$ of finding a galaxy with circular velocities $v_c$ in a
DM halo of circular velocity $V_c$.  Then the total number density of
galaxies with circular velocity $v_c$ (and the corresponding
probability density $p(v_c,t)$) is computed from
$N_g(v_c,t)=\int_{v_c}^{\infty}\,dV_c\,N(V_c,t)\,f(v_c,V_c,t)$, where
the halo distribution $N(V_c,t)$ takes the canonical Press \&
Schechter form.

Finally, for each $v_c$ we compute the probability distribution
$p(v_c',v_c,t)$ of having a galaxy at time $t$ with circular velocity
$v_c'$ included in a galaxy with circular velocity $v_c$ at time
$t+\Delta t$. This is given by
$p(v_c',v_c,t)=\int_{v_c'}^{\infty}\int_{v_c}^{\infty}
\,dV_c\,dV_c'\,P(V_c',V_c,t)\,f(v_c',V_c',t)\,f(v_c,V_c,t)$; 
from this we also derive the associated conditional
probability $p(v_c',t|v_c)=p(v_c',t|v_c)/p(v_c,t)$ that, {\it given} a
galaxy with velocity $v_c$ at time $t$, a galaxy $v_c'$ has been
included in it at $t-\Delta t$.

The average increments of the cold gas mass and the stellar mass in
galaxies with circular velocity $v_c$ are given by the equations in
White \& Frenk (1991); in differential form these read
\begin{equation}
m_c(v_c,t+\Delta t)=m_c(v_c,t) + \Delta t\,\int_0^{v_c}\,dv_c'
\,{N_g(v_c',t)\over N_g(v_c,t)}\,p(v_c',t|v_c)\,m_c(v_c',t)
\end{equation}
\begin{equation}
m_*(v_c,t+\Delta t)=m_*(v_c,t) + \Delta t\,\int_0^{v_c}\,dv_c'
\,{N_g(v_c',t)\over N_g(v_c,t)}\,p(v_c',t|v_c)\,m_*(v_c',t)~.
\end{equation}
The hot gas reservoirs are assumed to be retained in the DM halos 
rather than in individual galaxies. The amount of hot gas in halos 
with velocity $V_c$ is given by 
\begin{equation}
m_h(V_c,t+\Delta t)=m_*(V_c,t) + \Delta t\,\int_0^{V_c}\,dV_c'
\,{N(V_c',t)\over N(V_c,t)}\,
{\partial P(V_c',t|V_c)\over \partial V_c'\partial t}\,m_h(V_c',t)~.
\end{equation}

{\bf c)} The mass of the cold phase is further incremented by the cooling:
the gas that cools in a halo with $V_c$ in the time step $\Delta t$ is
$\Delta m_c(V_c)= 4\pi\,r_{cool}^2\,\rho_g(r_{cool})\,\Delta
r_{cool}$, where $\rho_g$ is the gas density and the dependencies of
the cooling radius $r_{cool}$ on $V_c$ and $t$ are given, e.g., in
Somerville \& Primack (1998). Such cooled gas is assigned to the
galaxies which have the same circular velocities of their DM halo, so
that the average hot gas in galaxies with circular velocity $v_c$ is
$\Delta m_c(v_c)=
\Delta m_c(V_c)\,N_g(V_c,V_c)/\int dV_c'\,N_g(V_c,V_c')$.

The amount of cold gas which forms stars is then computed and the
corresponding mass is transferred from the cold phase to the star
phase. Such amount is $\Delta m_*(v_c,t) = \dot M_*\,\Delta t$ and it
is regulated by the star formation rate $\dot
m_*(v_c,t)=\big[m_c(v_c,t)-m_*(v_c,t)-m_h(v_c,t)\big]/\tau_*$, which
as in Cole et al. (1994).  The timescale
$\tau_*(v_c)=\tau_*^0\,(v_c/300~{\rm km/s})^{\alpha_*}$ is
parametrized according to Cole et al. (1994).

A final transfer from the cold phase back to the hot phase, due to
supernovae feedback, involves the amount $\Delta m_h(v_c,t) = \beta(v_c)
\,\Delta m_*(v_c,t)$ where $\beta = (v_c/v_h)^{\alpha_h}$ is defined
by the free parameters $v_h$ and $\alpha_h$.  The total hot gas
returned to a halo with velocity $V_c$ by the member galaxies is
$\Delta m_h(V_c)=\int_0^{V_c}\,dv_c\,N_g(v_c,V_c,t)\Delta m_h(v_c,t)$.

{\bf d)} For each circular velocity $v_c$, the associated luminosity at
wavelength $\lambda$ is computed by the convolution
\begin{equation} 
S_{\lambda} = (1/\Gamma)\,\int_0^t\,\phi_{\lambda}(t-t')\,m_*(t')\,dt'~,
\end{equation} 
where the spectral energy distribution of luminous stars
$\phi_{\lambda}(t)$ is taken from the stellar population synthesis
models by Bruzual \& Charlot (1996). The output normalization
depends on the free parameter $\Gamma$, i.e., the fraction of
``stellar'' mass locked in non-luminous objects (planet-like
objects). The dust absorption is included in the computation
with the same procedure described in Somerville \& Primack (1998).

{\bf f)} We increment the current time by $\Delta t$ and repeat the whole
procedure from step b) until the output time is reached.

As explained in the text, we choose our parameters in order to fit
both the local observed Tully-Fisher relation and the local observed
B-band luminosity function. For the star formation parameters, the
above procedure yields the following set:
\newline $\tau_*^0=2$ Gyr, $\alpha_*=-0.5$, $v_h=100$ km/s and
$\alpha_h$=5 for the stellar formation.\newline $\Gamma=1$ and
$\Omega_b=0.06$ for the normalization of the mass-to-luminosity
ratio. \newline $\tau_o=0.1$ dynamical times, and $\alpha_m=0.2$ for
the galaxy merging rate. In addition, a cutoff in the cooling for
halos with $v_c>500$ km/s is included (see Somerville \&
Primack 1998).

\end{appendix}

\newpage

\centerline {\bf FIGURE CAPTIONS}
\bigskip\bigskip

\noindent
Fig.~1 The photometric redshift distribution of galaxies with $I\leq
25$ in the NDF.

\bigskip
\noindent
Fig.~2 Deconvolved half light radii as a function of true values in simulated
data. Filled points are for the bright objects, empty for the faint ones. 
Error bars are one sigma confidence intervals. 

\bigskip
\noindent
Fig.~3 Half light radii obtained deconvolving the HST and NDF images
of the same objects in the I band. The stright line represents equal
values.

\bigskip
\noindent
Fig.~4 Distribution of the half light radius for galaxies with $I\leq
25$. The values are in log (arcsec).

\bigskip
\noindent
Fig.~5 Disk scale length ($r_d=r_{hl}/1.68$) as a function of
luminosity for galaxies with $I\leq 25$ in two redshift
intervals. Filled circles correspond to the disk-dominated subsample
whose selection is shown in Fig.~6 and discussed in Sect.~4.

\bigskip
\noindent 
Fig.~6 Histogram of the differential size distribution for two
redshift intervals. The plotted numbers are galaxies per 5 arcmin$^2$
field per kpc.

\bigskip
\noindent
Fig.~7 Distribution of the galaxies with $I\leq 25$ in the asymmetry -
concentration plane (see text). The two straight lines delimit regions
corresponding to different spectral types according to Abraham et
al. (1996). The region inside the lines is occupied by normal spirals.
The outside region towards low concentration and high asymmetry is
occupied by galaxies with irregular morphology. Finally, the outside
region towards high concentration and lower asymmetry is occupied
by S0 and few elliptical galaxies.

\bigskip
\noindent
Fig.~8 Size distribution of galaxies as shown in Fig.~6 but for
galaxies having compactness and asymmetry typical of normal spirals.
These galaxies are inside the region delimited by the two straight
lines in Fig.~7. The continuous curve is the distribution predicted by
the CDM model described in Sect.~5. To compare with the observed data in
our sample, we have considered in the predicted counts a magnitude
limit at $I<25$. This limit has the effect of selecting brighter (and
hence larger) objects at higher $z$. Such an effect tends to balance the
decrease with $z$ of the disk size predicted by CDM models (as shown
in fig. 10 at given absolute magnitude); thus the predicted distributions 
in the two redshift intervals show an apparent lack
of evolution with $z$. 

\bigskip
\noindent
Fig.~9 {\it Top:} Tully-Fisher relation; the shaded area represents
the region allowed by the observations; the continuous curve is the
predicted T-F relation for the dark matter circular velocity while the
dashed curve is the T-F relation for the disk circular velocity. {\it
Middle:} the predicted local luminosity function; the data are adapted
from Zucca et al. (1998). {\it Bottom:} the predicted total star
formation rate as a function of redshift.

\bigskip
\noindent
Fig.~10 The fractional size distribution of galaxies predicted by the
CDM model at different $z$ for two ranges of absolute magnitude. The
distribution results from the distribution of angular momentum
$\lambda$, which is related to the disk radius $R_d$ and circular
velocity $v_d$ by eq. 5. The dependence on the circular velocity is
translated into a dependence on absolute magnitude through the
Tully-Fisher relation predicted by the model and shown in the first
panel of Fig.~9. Note the growth of the average disk size with cosmic
time typical of CDM scenarios.

\bigskip
\noindent
Fig.~11 Distribution of galaxies in the luminosity-size plane in four
redshift intervals. The disk radii are in kpc.  The shaded region
represents the region allowed by the model. The upper and lower lines
correspond to the 90\% and 10\% points of the angular momentum
distribution.  The solid line corresponds to the 50\% point of the
same distribution (see the text for details).


\begin{references}

\reference{}{Abraham, R. G., Tanvir, N. R., Santiago, B., Ellis, R. S., 
Glazebrook, K., van den Berg, S. 1996, MNRAS, 279, L47}

\reference{}{Arnouts, S., D'Odorico, S., Cristiani, S., Zaggia, S., Fontana, 
A., \& Giallongo, E. 1999, A\&A, 341, 641}

\reference{}{Baugh, C.M., Cole, S., Frenk, C.S., \& C.G. Lacey 1998, ApJ, 
498, 504}

\reference{}{Bendinelli, O. 1991, ApJ, 366, 599}

\reference{}{Bertin, E., \& Arnouts, S.  1996, A\&AS, 117, 393} 

\reference{}{Bond, J.R., Cole, S., Efstathiou, G., \& Kaiser, N. 1991, ApJ
379, 440}

\reference{}{Bouwens, R., Broadhurst, T. J., Silk, J. 1997, ApJ, 506, 579}

\reference{}{Bower, R. 1991, MNRAS, 248, 332}

\reference{}{Brinchmann, J., et al. 1998, ApJ, 499, 112}

\reference{}{Bruzual, A. G., \& Charlot, S. 1993, ApJ, 405, 538}

\reference{}{Cole, S., \& Lacey, C.G. 1996, MNRAS, 281, 1176}

\reference{}{Cole, S., Aragon-Salamanca, A., Frenk, C.S., Navarro, J.F., \& 
 Zepf, S.E. 1994, MNRAS, 271, 781}

\reference{}{Connolly, A. J., Szalay, A. S., Dickinson, M., Subbarao, M. U.,
\& Brunner, R. J. 1997, ApJ, 486, L11}


\reference {}{Cowie, L. L., Songaila, A., Hu, E., \& Cohen, J. G.  1996, AJ,
112, 839}

\reference{}{Driver, S. P., Fernandez-Soto, A., Couch, W. J., et al. 1998, 
ApJ, 496, 93}


\reference {}{Ellis, R. S., Colless, M., Broadhurst, T. J., Heyl, J. S.,
Glazebrook, K. 1996, MNRAS, 280, 235}

\reference{}{Fasano, G., Cristiani, S., Arnouts, S., Filippi, M. 1998, AJ, 115,
1400}

\reference{}{Fontana, A., et al. 1999, AJ, submitted}

\reference {}{Giallongo, E., D'Odorico, S., Fontana, A., Cristiani, S., 
Egami, E., Hu, E., McMahon, R. G.  1998, AJ, 115, 2169}

\reference {}{Glazebrook, K., Abraham, R., Santiago, B., Ellis, R., \&
	Griffiths, R. 1998, MNRAS, 297, 885}

\reference{}{Kauffmann, G., White, S.D.M., , and Guiderdoni, B. 1993, MNRAS, 
  264, 201}

\reference{}{Lacey, C., \& Cole, S. 1993, \mnras, 262, 627}

\reference{}{Lilly, S. J., Tresse, L., Hammer, F., Crampton, D., \& 
Le F\`evre, O. 1995, ApJ, 455, 108}

\reference{}{Lilly, S. J., et al. 1998, ApJ, 500, 75}

\reference{}{Madau, P., Ferguson, H. C. L., Dickinson, M., Giavalisco, M.,
Steidel, C. C., Fruchter, A., 1996, MNRAS, 283, 1388}

\reference{}{Mo, H.J, Mao S., \& White, S.D.M. 1998, MNRAS, 295, 319} 

\reference{}{Navarro, J.F., Benz, W. 1991, ApJ, 380, 320}

\reference{}{Navarro, J.F., Steinmetz, M. 1997, ApJ, 478, 13}

\reference{}{Navarro, J.F., White, S.D.M. 1997, MNRAS, 267, 401}

\reference{}{Navarro, J.F., Frenk, C.S., \& White, S.D.M. 1997, 
ApJ, 490, 493}

\reference{}{Odewahn, S. C. et al. 1996, ApJ, 476, L13} 

\reference{}{Pascarelle, S. M., Lanzetta, K. M., \& Fernandez-Soto, A. 1998,
preprint, astro-ph/9809295}

\reference {}{Primack, J.R., Holtzman, J., Klypin, A., \& Caldwell, D.O. 1995, 
 Phys. Rev. Lett., 74, 2160}

\reference{}{Roche, N., Ratnatunga, K., Griffiths, R. E., Im, M., \& Naim, A.
 1998, MNRAS, 293, 157}

\reference{}{Schade, D., Lilly, S. J., Crampton, D., Hammer, F., Le F\`evre, 
O., Tresse, L. 1995, ApJ, 451, L1}

\reference{}{Schade, D., Lilly, S. J., Le F\`evre, O., Hammer, F., Crampton, D.
1996, ApJ, 464, 79}


\reference {}{Somerville, R., \& Primack, J.R. 1998, preprint,
astro-ph/9802268}

\reference {}{Steidel, C. C., Adelberger, K. L., Giavalisco, M., Dickinson,
M., \& Pettini, M. 1999, ApJ in press, astro-ph/9811399}

\reference{}{Steinmetz, M., \& Bartelmann, M., 1995, MNRAS, 272, 570}

\reference{}{Warren, M.S., Quinn, P.J., Salomon, J.K, \& Zurek, W.H. 1992, 
ApJ, 399, 405}

\reference{}{White, S.D.M.,  \& Frenk, C.S. 1991, ApJ, 379, 52}


\reference{}{Williams, R. E., et al. 1996, AJ, 112, 1335}

\reference{}{Zucca, E.,  et al. 1997, A\&A, 326, 477}

\end{references}
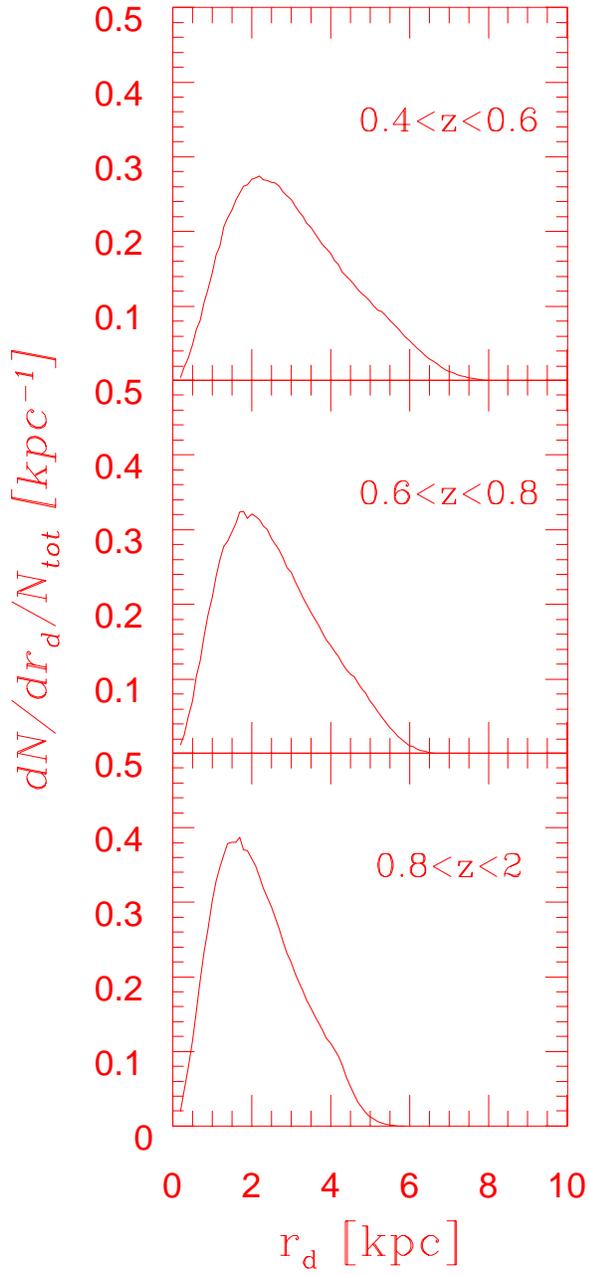
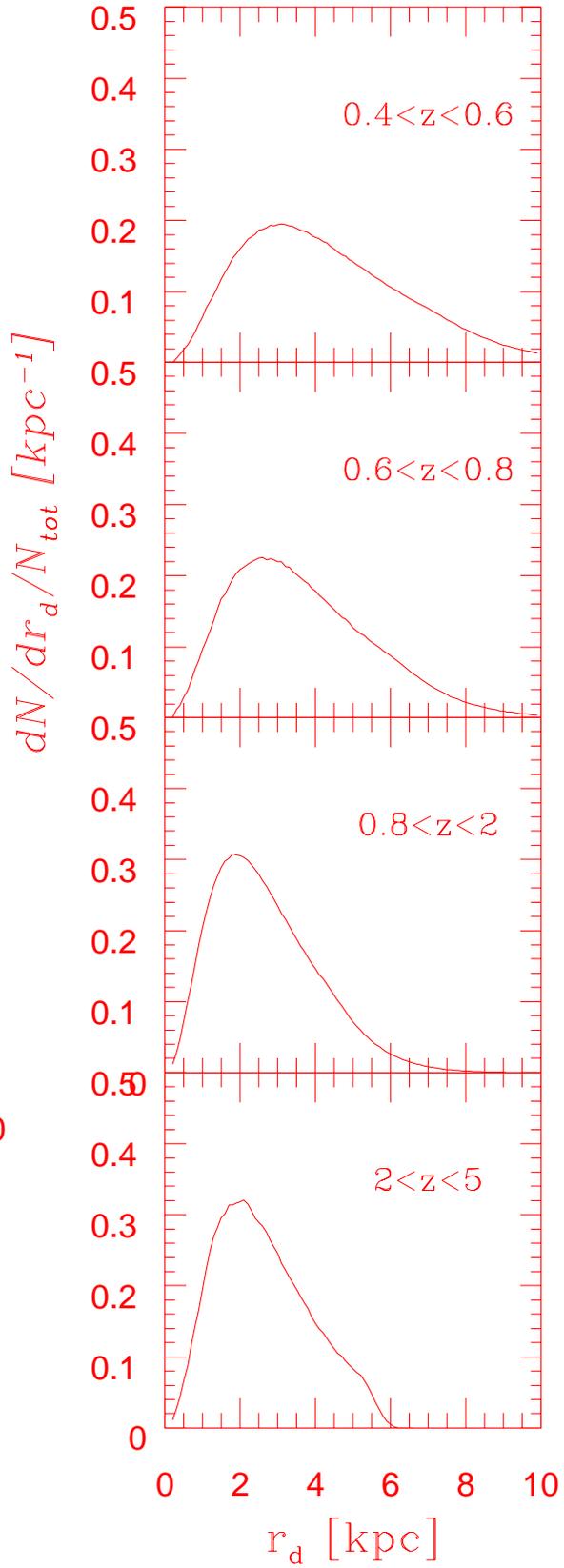
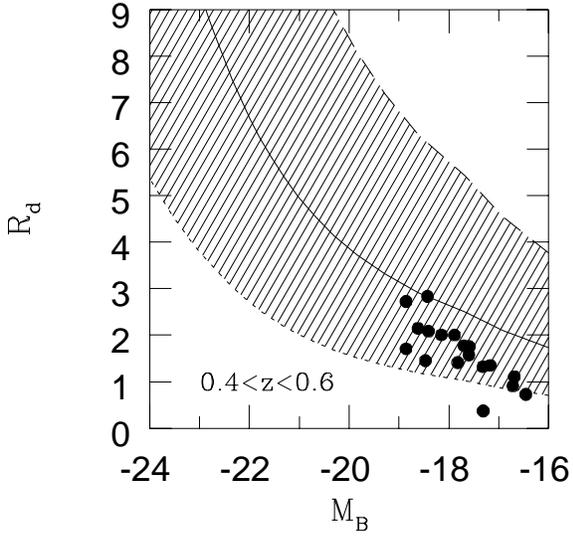
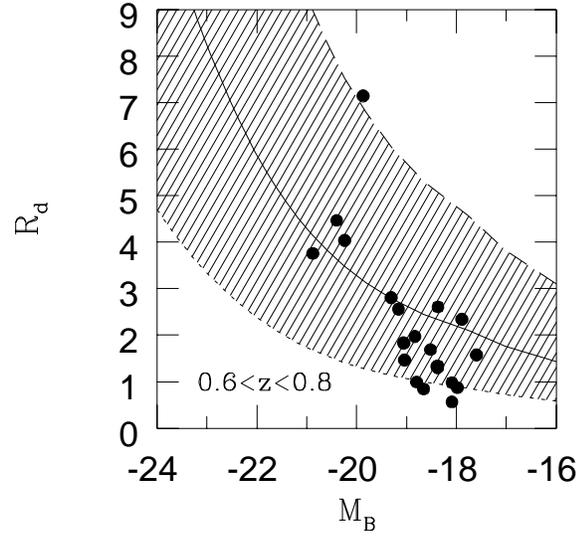
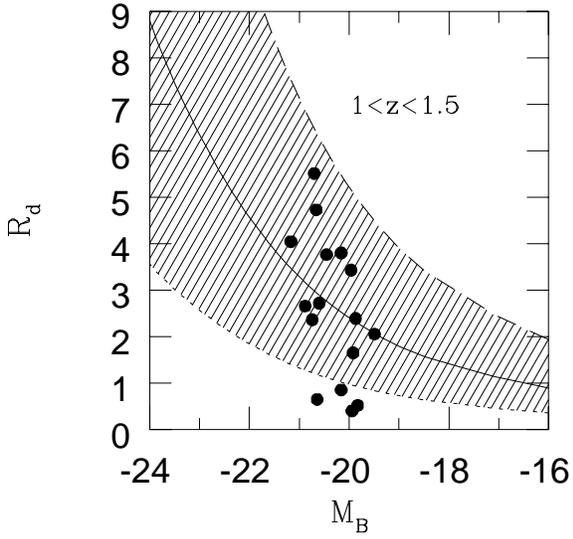
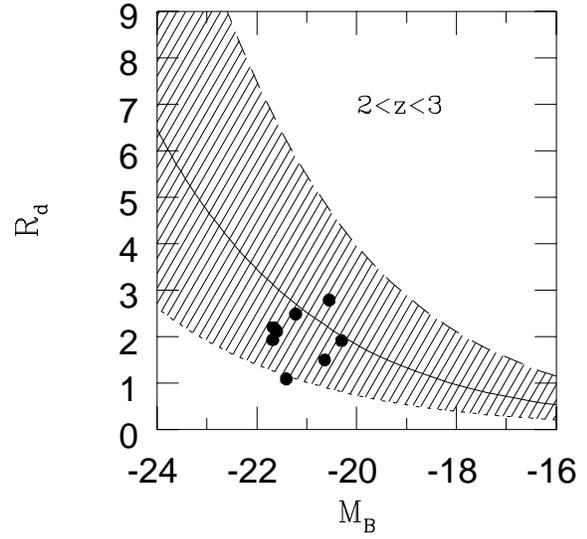
\end{document}